# Broadband silicon polarization beam splitter based on Floquet engineering

**NING MA[1], YU CHEN[1], YUNJIE LI[1], SHUBIN HUANG[2], WEN-DI LI[2], MINGHUA CHEN[3], AND CIYUAN QIU[1,*]**

[1]*State Key Laboratory of Advanced Optical Communication Systems and Networks, Department of Electronic Engineering, Shanghai Jiao Tong University, Shanghai, China*

[2]*Department of Mechanical Engineering, The University of Hong Kong, Hong Kong, SAR, 999077, China*

[3]*Department of Electronic Engineering, Tsinghua University, Beijing 100084, China*

**A broadband silicon polarization beam splitter (PBS) is proposed and experimentally demonstrated based on Floquet-engineered directional couplers. The total length of the coupling structure is 20 μm. By periodically modulating the waveguide widths of the directional couplers, the power exchange between the two waveguides for the transverse-electric (TE) mode is suppressed, while the power coupling for the transverse-magnetic (TM) mode is enhanced. The fabricated PBS exhibits polarization extinction ratios (PERs) > 20 dB for both polarizations over a broad wavelength range of 1483 nm–1620 nm. Additionally, the measured insertion losses (ILs) are 0.15 dB and 1.2 dB at 1550 nm for TE and TM polarizations, respectively.**

Silicon photonics has become a promising solution for data interconnects owing to its low power consumption and high bandwidth, particularly in the era of artificial intelligence (AI) [1, 2]. To further scale the transmission capacity, multiplexing technologies are in high demand, in which polarization-division multiplexing (PDM) has been recognized as an effective approach. In such PDM systems, polarization beam splitters (PBSs) serve as essential components, enabling the separation of TE and TM modes with low insertion losses (ILs) and high polarization extinction ratios (PERs).

Over the past decades, several PBS structures have been demonstrated on silicon photonic platforms, including multimode interferometers (MMIs) [3-6], directional couplers (DCs) [7-12], and subwavelength grating (SWG) [3, 4, 10, 13] assisted structures. Among them, DC-based PBSs show excellent performance in terms of compact footprint, high PERs, and fabrication compatibility. However, conventional DC-based PBSs generally exhibit strong wavelength dependence, which limits their operational bandwidth and robustness [7, 8].

Recently, Floquet engineering has been introduced into integrated photonics as an effective theoretical framework for analyzing and designing periodically modulated photonic structures [14-21]. By applying periodic perturbations along the propagation direction, such systems enable additional control over the light propagation and mode coupling. This approach has been implemented through various forms of geometric modulation, including waveguide width variation[15, 17, 21] and sinusoidal bending[16, 18-20]. These modulation schemes provide new degrees of freedom for photonic device design.

In this Letter, we propose and experimentally demonstrate a broadband silicon PBS based on Floquet-engineered directional couplers. By introducing a sinusoidal modulation of the waveguide width, the proposed device achieves polarization-dependent coupling for the TE and TM polarizations, allowing efficient polarization separation in a short coupling length. Numerical simulations indicate that, for the TM polarization, the proposed PBS achieves polarization extinction ratios (PERs) > 20 dB over a 155-nm wavelength bandwidth from 1480 nm to 1635 nm, with less than 1-dB insertion losses. For the TE polarization, the PBS exhibits PERs > 20 dB over a broad wavelength range from 1450 nm to 1650 nm, accompanied by insertion losses below 0.1 dB. For the fabricated device, experimental results demonstrate PERs > 20 dB for both polarizations over the wavelength range between 1483 nm and 1620 nm, with corresponding insertion losses of 0.15 dB for TE and 1.2 dB for TM at 1550 nm, respectively.

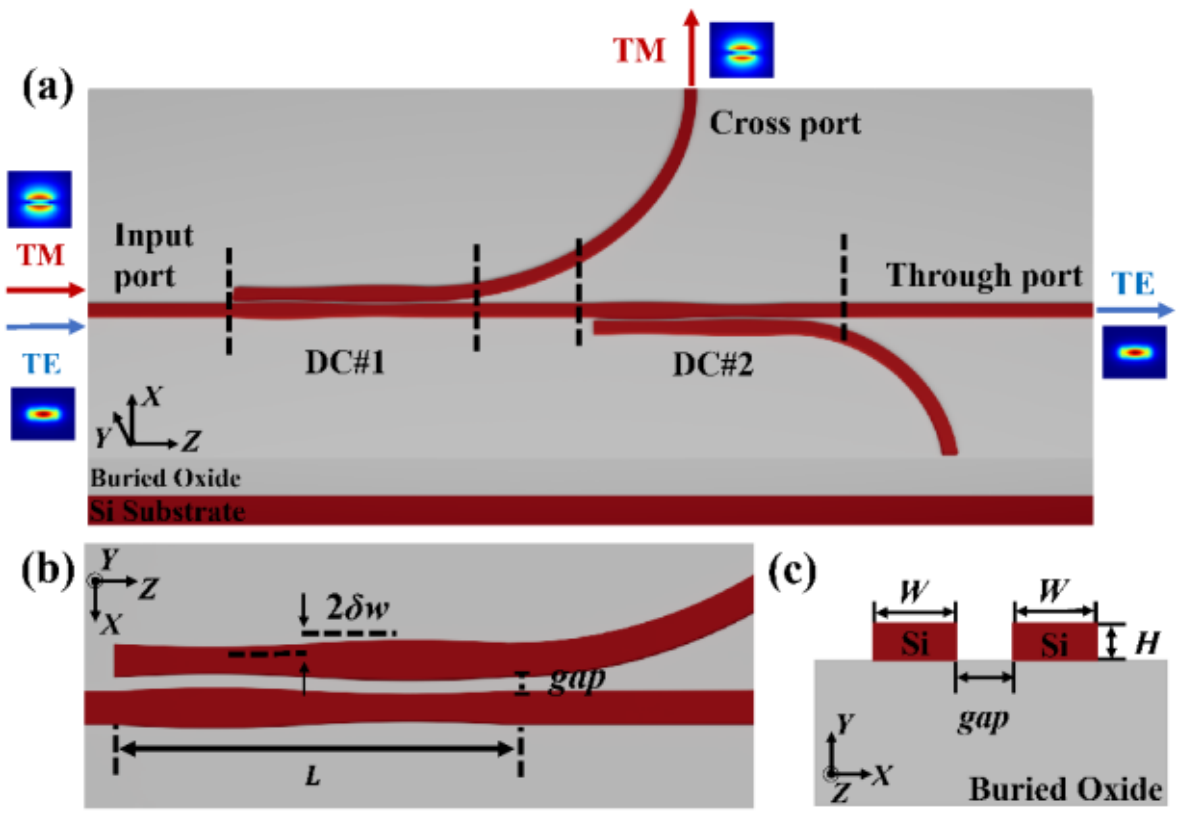

**Fig.1** The schematic configuration of the proposed PBS. (a) 3D view (b) Top view of the coupling region (c) Cross-section view.

---

Figure 1 shows the schematic layout of the proposed PBS, which is implemented on a silicon-on-insulator (SOI) platform. The three-dimensional view of the device is shown in Fig. 1(a), which consists of two identical cascaded directional couplers (DC#1, DC#2). Here, each DC consists of a pair of parallel strip waveguides, whose widths are sinusoidally modulated along the propagation direction. When light is launched into the input port, the fundamental TE mode remains highly confined to the input waveguide and propagates to the through port. In contrast, the power of the fundamental TM mode is efficiently transferred to the cross port by the PBS. The detailed geometry of the PBS is presented in Figs. 1(b) and 1(c). The top view in Fig. 1(b) illustrates a directional coupler that has its width sinusoidally modulated with a period $L$ and an amplitude $\delta w$. Meanwhile, as illustrated in Fig. 1(c), the device consists of two identical coupler strip waveguides with a height $H$ and an average width $W$, separated by a gap.

To understand the operating principle, the optical behavior of each DC can be modeled as a Floquet-driven system with spatial modulation frequency $\Omega = 2\pi/L$ along the propagation axis $z$, where $L$ is the modulation period of the waveguides as shown in Fig. 1(a). Under the paraxial approximation, for each polarization, the light evolution in such a coupled system is governed by a Schrödinger-like equation [15, 21]:

$$i\frac{\partial}{\partial z}|\psi(z)\rangle = H(z)|\psi(z)\rangle \tag{1}$$

where $H(z)$ denotes the $z$-dependent Hamiltonian of the system. For weak width modulation, the coupling coefficient between the waveguides can be taken as a constant $\kappa$. The Hamiltonian can thus be expressed as:

$$\boldsymbol{H(z)} = \sum_{m=1}^{2} \boldsymbol{\beta_0 \alpha_m^\dagger \alpha_m} + \sum_{m=1}^{2} \boldsymbol{A \sin(\Omega z + \phi_m)\, \alpha_m^\dagger \alpha_m}$$

$$+\boldsymbol{\kappa(\alpha_1^\dagger \alpha_2 + \alpha_1 \alpha_2^\dagger) + h.c.}, \tag{2}$$

where $\beta_0$ is the propagation constant, $a_m^\dagger(a_m)$ is the creation (annihilation) operator, $A = 2\pi n'_{eff}/\lambda$ is the modulation amplitude determined by the effective index detuning $n'_{eff}$, $\phi_m$ represents the modulation phase of the $m$-th waveguide, and $h.c.$ denotes the Hermitian conjugate. Note that the above-mentioned parameters, including $\beta_0$, $\kappa$, and $n'_{eff}$, are polarization-dependent due to the different modal dispersions of the TE and TM modes.

In Floquet theory, such a periodically driven system is characterized by a quasi-energy $\varepsilon$, which can be obtained from equation (2) [15]. When the modulation phase difference between adjacent waveguides is set to $\pi$ ($\phi_2 - \phi_1 = \pi$), the quasi-energy collapse occurs, resulting in strong field localization [15]. Then, the periodically driven system can be described by an effective coupling, which is expressed as[15]:

$$\boldsymbol{\kappa_{eff} = \kappa J_0(\xi)} \tag{3}$$

where $J_0(\xi)$ is the zero-order Bessel function of the first kind, $\xi = 4\pi n'_{eff}/(\lambda\Omega)$ denotes the normalized modulation strength and $\lambda$ is the operating wavelength. Thus, if $\xi$ is set to the roots of the function ($\xi_0 = 2.405, 5.520, \ldots$), the coupling between waveguides can be completely suppressed, which is the theoretical zero-coupling condition.

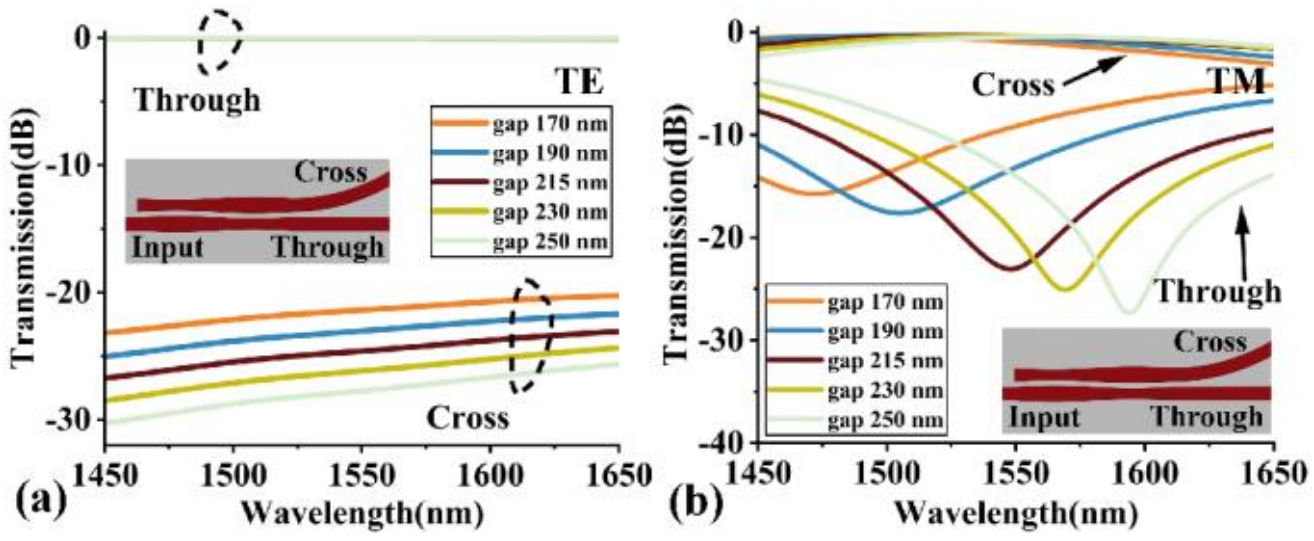

**Fig.2** Simulated transmission spectra of DC#1 for (a) the TE mode and (b) TM mode with different waveguide gaps and fixed modulation parameters ($\delta w = 50$ nm and $L = 7$ μm).

Next, structural parameters are properly chosen to achieve enhanced polarization beam splitter performance. Here, to suppress the coupling of the TE polarization, one needs to obtain the above-mentioned theoretical zero-coupling condition for the TE polarization. Thus, the modulation parameter $\delta w$ and the period length $L$ are set to 50 nm and 7 μm, respectively. Then, to enhance the TM coupling, the proper gap between the two waveguides is required. Figures 2(a) and 2(b) show the simulated transmission spectra of DC#1 for the TE and TM polarizations, respectively, with different waveguide gaps. It can be observed that the TE polarization remains effectively suppressed over the entire range of gap values, indicating that the selected modulation parameters satisfy the TE-mode suppression condition. Meanwhile, for the TM polarization, light is efficiently coupled to the cross port. Moreover, at the through port, there is some residual light power for the TM mode, while the transmission varies with the waveguide gap. Based on the trade-off between coupling efficiency and bandwidth, a gap of 215 nm is chosen, providing a bandwidth of about 35 nm with PERs > 20 dB for both TE and TM polarizations.

To further improve the bandwidth, a cascaded configuration consisting of two identical directional couplers is adopted. As illustrated in Fig. 1(a), the second directional coupler (DC#2) is cascaded at the through port of DC#1. Thus, the residual TM power at the through port can be further filtered, enhancing the bandwidth and the PERs of the device. Figures 3(a) and 3(b) show the simulated optical field distributions at 1550 nm for the TE and TM polarizations, respectively. The simulated results indicate that the PBS can effectively separate the TE and TM modes.

Figures 3(c) and 3(d) present the simulated transmission spectra of the TE and TM polarization modes for the cascaded PBS structure at the through port and cross port. For TE polarization, the PBS exhibits PERs > 20 dB in a broad band of 1450 nm-1650 nm, with insertion losses < 0.1 dB, as shown in Fig. 3(c). From Fig. 3(d), the proposed PBS has PERs > 20 dB in the wavelength range of 1480 nm-1635 nm, with insertion losses < 1 dB for TM polarization. Note that the bandwidth for TM polarization is broadened from 35 nm to 155 nm by using cascaded structures.

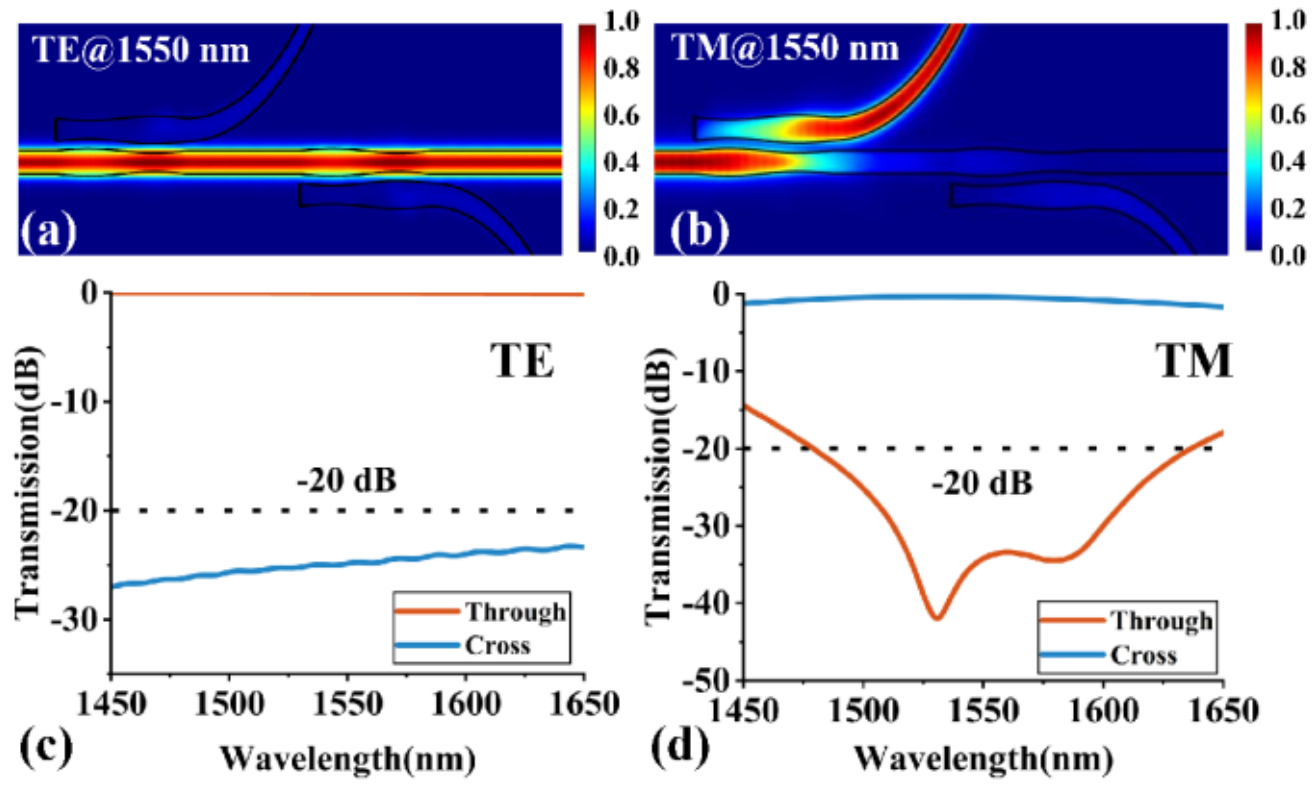

**Fig. 3** Simulated normalized optical field magnitude of the cascaded PBS for (a) the TE polarization and (b) the TM polarization. Simulated transmission of the cascaded PBS for (c) the TE polarization and (d) the TM polarization. In (a) and (b), the black outlines indicate the waveguide boundaries, while the color maps represent the normalized field magnitude.

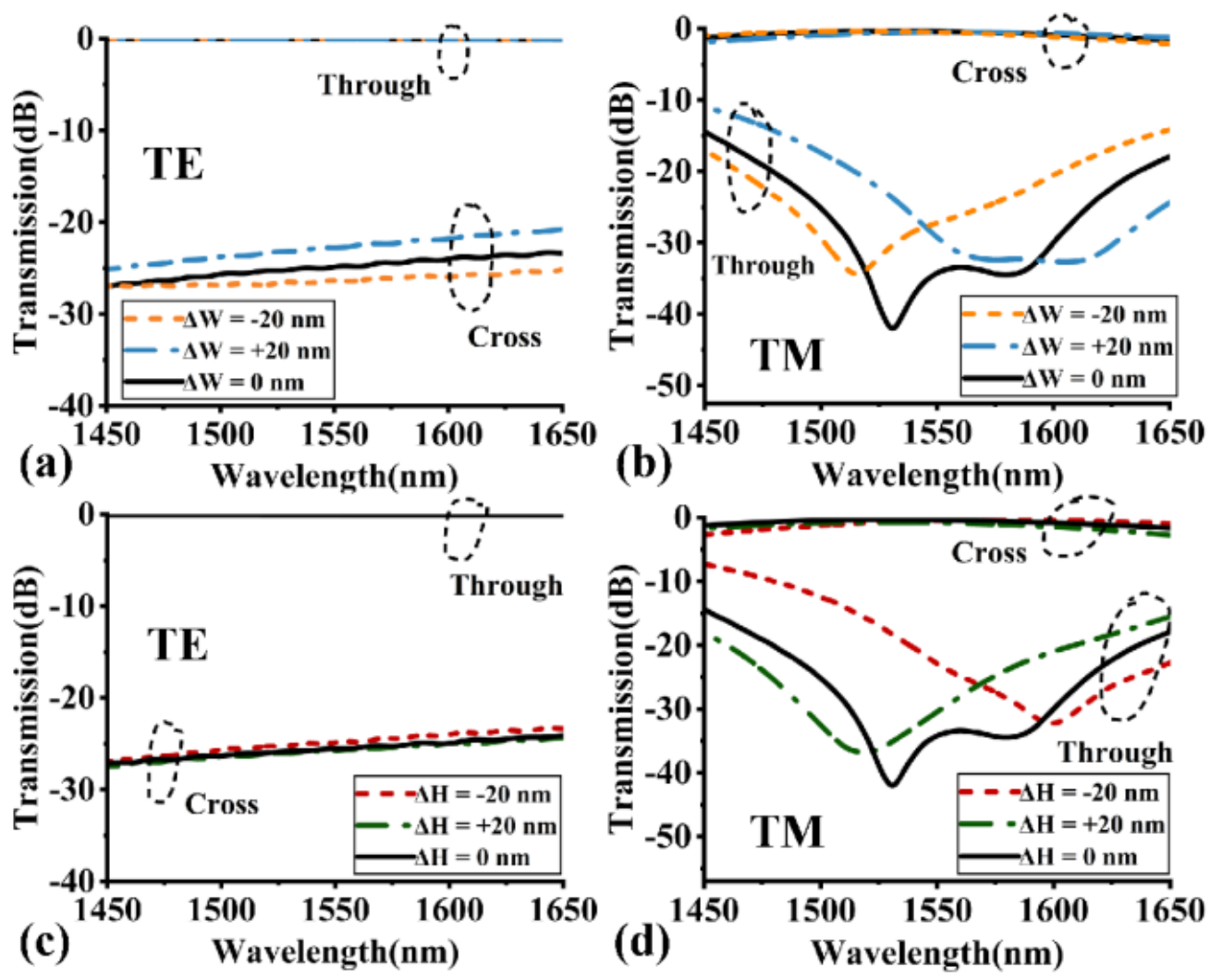

**Fig. 4** Fabrication tolerance analyses of the proposed PBS. (a) and (b) show the influence of waveguide width variations ($\Delta W$) on the transmission spectra for TE and TM modes, respectively. (c) and (d) illustrate the impact of waveguide height variations ($\Delta H$) for TE and TM modes, respectively.

To evaluate the practical feasibility of the proposed PBS, a fabrication tolerance analysis is performed, considering the variations of both waveguide width ($\Delta W$) and height ($\Delta H$). Figures 4(a)–4(d) present the simulated transmission spectra of the TE and TM modes under different values of $\Delta W$ and $\Delta H$. One can find that the PBS maintains PERs > 20 dB over a bandwidth of > 100 nm for variations up to ±20 nm, demonstrating robust tolerance against fabrication errors.

The proposed PBS was then fabricated on a SOI (Silicon-on-insulator) wafer using E-beam lithography (EBL, Vistec EBPG 5200+) and inductively coupled plasma dry-etching (ICP, SPTS). The SOI wafer consists of a 220-nm-thick silicon device layer and a 3-μm-thick $SiO_2$ buried oxide layer, with air serving as the upper cladding. Figure 5(a) shows the optical microscope image of the fabricated device. And, the scanning electron microscope (SEM) view is shown in Fig. 5(b), where the sinusoidally modulated waveguide structure can be clearly observed. Notably, both TE- and TM-type grating couplers (GCs) are fabricated in this work. The TE-type grating has a period of 630 nm with a duty cycle of 0.5, while the TM-type grating has a period of 980 nm and a duty cycle of 0.6. Both couplers are etched with a shallow depth of 70 nm. The coupling losses of the grating couplers are approximately 7 dB/facet for the TE polarization and 7.5 dB/facet for the TM polarization at 1550 nm.

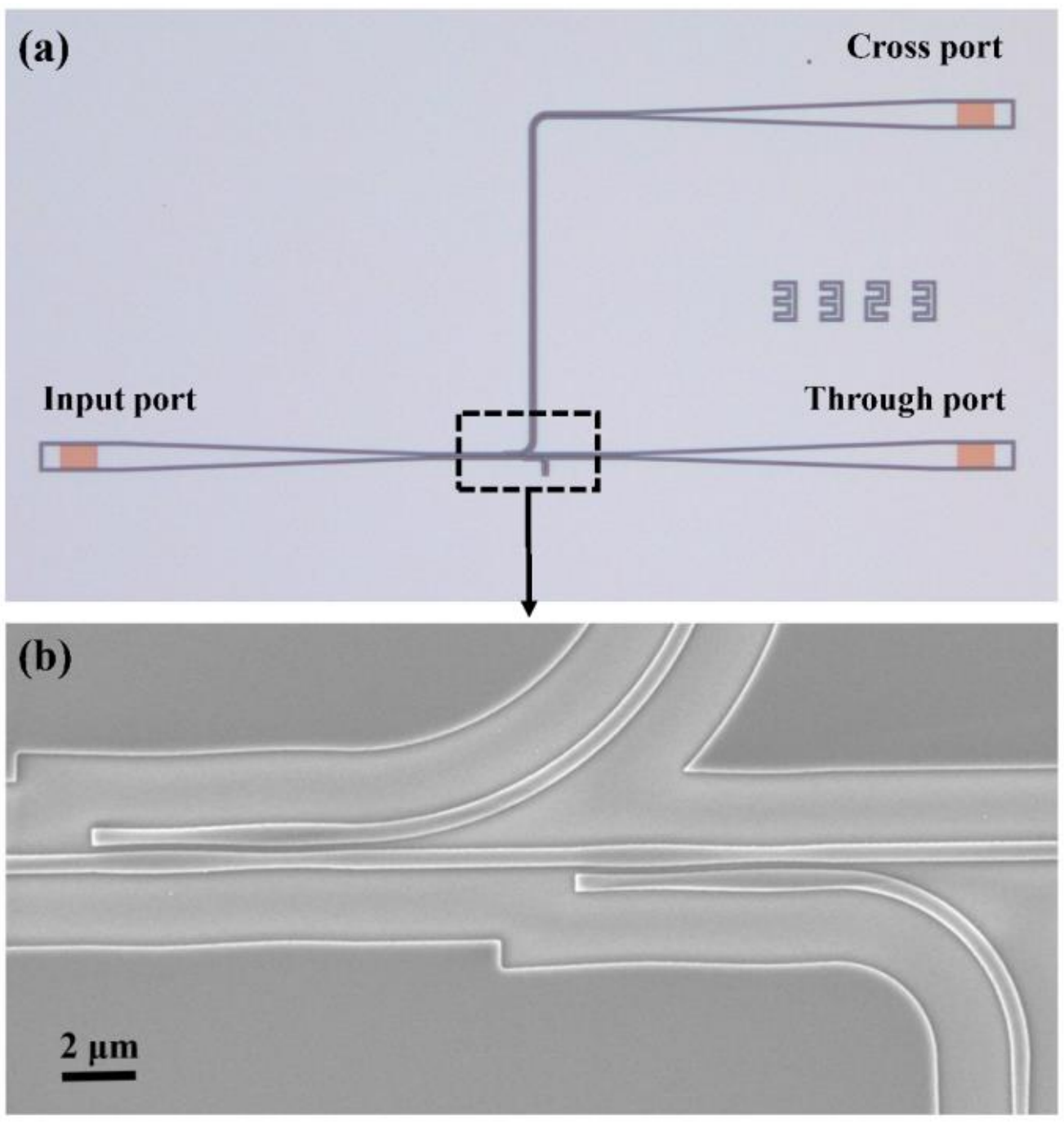

**Fig. 5** (a) Optical microscope image of the fabricated device. (b) SEM image of the coupling region

A tunable continuous wave laser (Santec TSL-570) and a power meter (Santec MPM-220) are used to measure the transmission spectra at wavelengths ranging from 1483 nm to 1637 nm. In the experiment, the laser power is set to be 0 dBm, and the polarization is adjusted by a polarization controller (PC). The light is injected into the device via the grating coupler. After passing through the device, the optical spectra of the through port and cross port are measured by the optical power meter, whose minimum detectable power is -80 dBm. The measured spectrum is then normalized with respect to the straight waveguide on the same chip with the same grating couplers.

Figures 6(a) and 6(b) show the normalized transmission spectra at the through ports and cross ports for TE and TM polarizations, respectively. It can be observed that for both polarizations, PERs > 20 dB can be achieved over the entire wavelength range from 1483 nm to 1620 nm, indicating broadband polarization splitting performance. In addition, the insertion losses for the TE polarization remain < 1 dB across the measured bandwidth, while the ILs for the TM polarization are < 2 dB. At 1550 nm, the insertion losses for the TE and TM polarizations are approximately 0.15 dB and 1.2 dB, respectively. Note that the performance of the fabricated PBS can be further improved through optimized experimental instruments and fabrication processes.

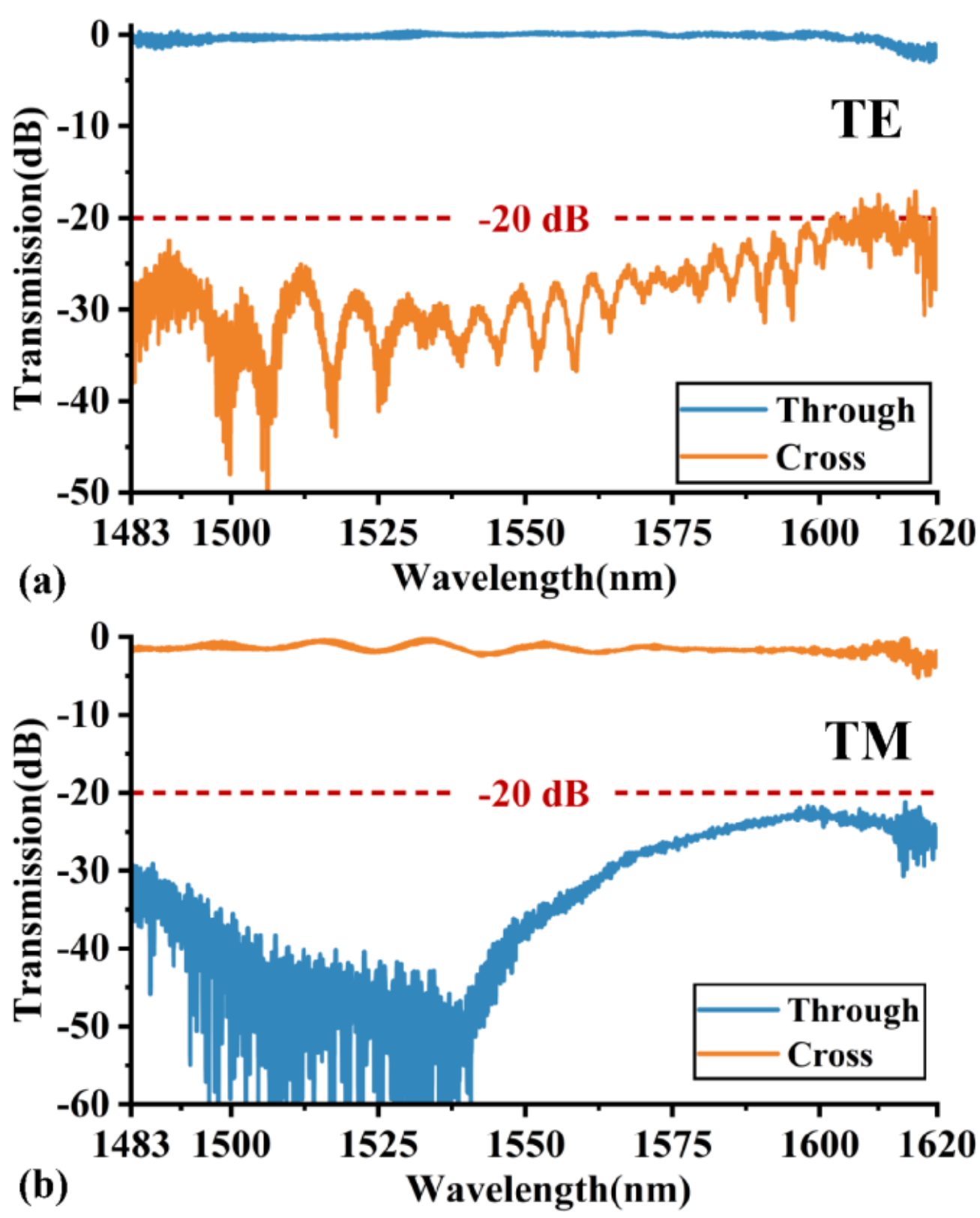

**Fig. 6** Normalized transmission spectra of the fabricated device for (a) the TE polarization and (b) the TM polarization.

In conclusion, a broadband PBS based on a Floquet-engineered directional coupler has been proposed and experimentally demonstrated. Sinusoidal width modulation is introduced into the waveguide. By designing the modulation amplitude and period, the power exchange for the TE polarization can be greatly suppressed while efficient power transfer is achieved for the TM polarization. The cascaded coupling structure has a total length of 20 μm. For the fabricated device, the measured bandwidths for PERs > 20 dB are ~137 nm, spanning from 1483 nm to 1620 nm. Additionally, the measured ILs are < 1dB for TE mode and < 2dB for TM mode. At 1550 nm, the insertion losses for the TE and TM polarization are approximately 0.15 dB and 1.2 dB, respectively. Therefore, the proposed PBS provides a broadband and fabrication-friendly solution for polarization division multiplexing in silicon photonic integrated circuits.

**Funding.** National Key Research and Development Program of China (2024YFE0204000); National Natural Science Foundation of China (62275149, U23A20356); Open Fund of State Key Laboratory of Optoelectronic Materials and Devices (SKLOEMD-OP-2025-G01).

**Acknowledgment.** The authors would like to thank the Center for Advanced Electronic Materials and Devices (AEMD) of Shanghai Jiao Tong University for the support in device fabrication and characterization.